\documentclass[12pt,reqno]{amsart}

\usepackage{amsmath}
\usepackage{amsfonts}
\usepackage{amsthm}
\usepackage{amstext}

\newcommand{\supp}{\operatorname{supp}}

\theoremstyle{plain}
\newtheorem{thm}{Theorem}[section]
\newtheorem{prop}[thm]{Proposition}
\newtheorem{lemma}[thm]{Lemma}
\newtheorem{cor}[thm]{Corollary}
\theoremstyle{definition}

\newtheorem*{acknowledgement}{Acknowledgement}

     \newtheorem{remark}[thm]{Remark}

     \numberwithin{equation}{section}

\title{Analyticity of the density of electronic wavefunctions}

\author[S. Fournais, M. and T. Hoffmann-Ostenhof, T. \O. S\o rensen]
{S. Fournais \and M. Hoffmann-Ostenhof \and T. Hoffmann-Ostenhof \and
T. \O stergaard S\o rensen}

\address[S. Fournais]{Laboratoire de Math\'{e}matiques\\
           Universit\'{e} Paris-Sud - B\^{a}t 425\\
           F-91405 Orsay Cedex\\ France.
           }
\email{fournais@imf.au.dk}

\address[T. \O stergaard S\o rensen]
{Department of Mathematical Sciences,
           Aalborg University,
           Fredrik Bajers Vej 7G,
           DK-9220 Aalborg East, Denmark.}
\email{sorensen@math.auc.dk}

\address[M. Hoffmann-Ostenhof]{Institut f\"{u}r Mathematik,
           Strudlhofgasse 4,
           Universit\"at Wien, A-1090 Vienna, Austria.}
\email{mhoffman@esi.ac.at}

\address[T. Hoffmann-Ostenhof]{Institut f\"ur Theoretische
Chemie, W\"ahringer\-strasse 17,
           Universit\"at Wien,
           A-1090 Vienna,
           Austria. }
\address[T. Hoffmann-Ostenhof, 2nd address]{
        The Erwin Schr\"{o}dinger International Institute for 
        Mathematical Physics,
              Boltzmanngasse 9,
              A-1090 Vienna, Austria.}
\email{thoffman@esi.ac.at}

\date{November 22, 2002}

\begin{document}

\thispagestyle{empty}

\begin{abstract} We prove that the electronic densities of
atomic and molecular eigenfunctions are real analytic in ${\mathbb R}^3$ 
away from the nuclei.
\end{abstract}

\maketitle



\section{Introduction and statement of the results}
We consider an $N$-electron molecule with $L$ fixed nuclei. The
non-relativistic Hamiltonian of the molecule is given by
\begin{multline}
   \label{Hmol}
   H = H_{N,L}(\mathbf R,\mathbf Z)=\sum_{j=1}^N\left({}-\Delta_j-\sum_{l=1}^L
   \frac{Z_l}{|x_j-R_l|}\right)\\
   +\sum_{1\le i<j\le N}\frac{1}{|x_i-x_j|}+ \sum_{1\le l<k\le
L}\frac{Z_lZ_k}
   {|R_l-R_k|},
\end{multline}
where $\mathbf R=(R_1,R_2,\dots ,R_L)\in\mathbb R^{3L}$, \(R_{l}\neq
R_{k}\)
for \(k\neq l\),
denote the positions of the $L$ nuclei
whose positive charges are given by $\mathbf Z=(Z_1, Z_2,\dots,Z_L)$.
The positions of the $N$ electrons are denoted by
${\bf x}=(x_1,x_2,\dots,x_N)\in\mathbb R^{3N}$, where $x_j$ denotes the 
position of the $j$'th electron in $\mathbb R^3$.
For shortness, we will sometimes write
\begin{equation}\label{H}
H = -\Delta + V({\bf x}),
\end{equation}
where $\Delta= \sum_{j=1}^N \Delta_j,$ is the $3N$-dimensional Laplacian, and
$V$ is the complete (many-body) potential. 
It is a standard fact that $H$ with domain $W^{2,2}(\mathbb R^{3N})$ is selfadjoint.

We consider eigenfunctions $\psi$ of $H$, i.e. solutions
$\psi \in L^2(\mathbb R^{3N})$ to the equation 
\begin{equation}\label{Hpsi}
H\psi=E\psi,
\end{equation}
with $E\in {\mathbb R}$.
Since we describe electronic wave functions, 
and the electrons are Fer\-mions, $\psi$ has to transform 
according to certain irreducible representations
of the symmetric group $\mathfrak S^N$. However, our results are independent 
of this condition and we do not impose it. 

Analyzing the spectrum of $H$ and calculating (usually by some approximation
scheme) the eigenvalues $E$  and the corresponding eigenfunction(s) $\psi$
 is the central theme of most of the investigations
done by quantum chemists and physicists. For the  
interpretation of these investigations the eigenfunction $\psi$ is much
too complex being a function of $3N$ variables and hence the 
{\bf one-electron density} $\hat\rho(x)$ plays a prominent role.
It is defined by 
\begin{equation}\label{rhohat}
\hat\rho(x)=\sum_{j=1}^N\int_{\mathbb R^{3N-3}}|\psi(\hat{\bf{x}}_j)|^2
d\hat{\bf{x}}_j
\end{equation}
where we use the notation 
$ \hat{\bf{x}}_j=(x_1,\dots,x_{j-1},x, x_{j+1},\dots, x_N)$
and $d\hat{\bf{x}}_j=
dx_1\dots dx_{j-1} dx_{j+1}\dots dx_N$.

The mathematical analysis of 
$H$ has mainly centered around the 
operator theoretical point of view, see for instance \cite{Kato},
\cite {ReSI} and references therein. 
The fact that \eqref{Hpsi} is an elliptic partial differential equation
has not been exploited in such depth; so many questions 
which are natural from a PDE point of view are not really understood.
In particular regularity questions concerning $\psi$ and $\hat \rho$ are
natural and interesting. 
Note first that $V$ is singular in
\begin{equation}\label{sing}
\Sigma=\bigg\{x\in\mathbb R^{3N}\,\big|\,
\big(\prod_{i=1}^N\prod_{\ell=1}^L|x_i-
R_\ell|\big)\big(\prod_{1\le i<j}^N|x_j-x_i|\big)=0\bigg\}
\end{equation}
and real analytic in $\mathbb R^{3N}\setminus \Sigma$.
Hence by standard methods of elliptic PDE, see for instance \cite{H0},
$\psi$ is real analytic in $\mathbb R^{3N}\setminus \Sigma$. 
The first results concerning the 
regularity of $\psi$ on all of $\mathbb R^{3N}$ are due to Kato 
\cite{Kato:1957}. He showed that 
$\psi$ is Lipschitz continuous 
and first formulated the well known cusp conditions,
which describe the behaviour of an eigenfunction near the points where two
particles are close to each other \cite[Theorems II and IIb]{Kato:1957}. See
 also 
the important paper by Simon \cite {Simon:1982} in which the Coulombic 
many particle potential $V$ is identified as a special member of the 
so called Kato class and some results 
concerning the regularity of solutions of Schr\"odinger equations 
(equations of type \eqref{Hpsi} with general $V$)
are given. 

Regularity results concerning the Coulombic case
extending Kato's result have been obtained more
recently in \cite{HHS:1994} and \cite{HHOS:2001};
see also the more complete references therein to other results concerning
regularity.

There are now two related problems:
\begin{itemize}
\item[i)] Describe in more detail how the specific structure of the 
singularities in $\Sigma$ turn up in the nonanalyticity of $\psi$.
Partial results can be found in the references cited above.
\item[ii)] Analyse the regularity properties of the 
one-electron density, defined in \eqref{rhohat},
an object which has an immediate  physical interpretation (see any 
textbook on quantum mechanics, for instance \cite{LandauLifschitz}) 
and enters all approximation schemes in a crucial way (Hartree-Fock, 
Thomas-Fermi, Density Functional Theories etc).
\end{itemize}
For the regularity questions concerning $\hat\rho$ defined in \eqref{rhohat}
it suffices to consider the 
(non-symmetrized) density $\rho$ defined by
\begin{align}
  \label{eq:rho}
  \rho(x) = \int_{{\mathbb R}^{3N-3}}
  \,|\psi(x,x_2,\ldots,x_N)|^2\,dx_2\cdots dx_N.
\end{align}

It is not clear {\it a priori} that $\rho$ is real analytic away from the nuclei since in \eqref{eq:rho} one integrates over subsets of $\Sigma$
where $\psi$ is not analytic. In two recent papers (\cite{FHHO} and 
\cite{FHHO1:2002})  
the present authors have 
shown that $\rho$ is smooth away from the positions of the nuclei (or in the 
case of an atom, away from the origin). 
The natural question is now whether $\rho$ is real analytic
away from the nuclei. This will be answered affirmatively in this
paper. Of course in the proof of this result new difficulties arise, in 
particular all the estimates have to be much more explicit.

\begin{thm}
\label{thm:density_analytic}
Let \(\psi \in L^2({\mathbb R}^{3N})\) satisfy the equation
$$
H\psi = E\psi,
$$
with $E\in {\mathbb R}$ and $H$ given by \eqref{Hmol}.
Let the density \(\rho\) be defined as in
\eqref{eq:rho}.\\
Then $\rho$ is a real analytic function in ${\mathbb R}^3
\setminus \{R_1,\ldots,R_L\}$.
\end{thm}

\begin{remark}[Atoms vs. molecules]
In order to keep notation simple, we will only give the proof of 
Theorem \ref{thm:density_analytic} in the case of an atom. In this 
case we only have one nucleus, which we place at the origin, 
so the potential $V$ is given by 
\begin{equation}\label{atomV}
V=-\sum_{i=1}^N\frac{Z}{|x_i|}+\sum_{1\le i<j}^N\frac{1}{|x_j-x_i|}.
\end{equation}
The necessary modifications for the molecular case
were indicated in the proof of the smoothness results 
in \cite{FHHO}. In the present proof of analyticity one 
has to make similar changes when working with molecules.
\end{remark}

\begin{remark}[Density matrices]
We get analogous results for the one electron density matrix $\gamma_1(x,x')$
and the $2$-electron density $\rho_2(x,x')$, 
which we will define in the following.
Let $(x,x') \in {\mathbb R}^6$. 
Let $\hat{\bf{x}}_j$ and $d\hat{\bf{x}}_j$ be as defined after \eqref{rhohat}, and define 
\begin{align*}
\hat{\bf{x}}'_j&=(x_1,\dots,x_{j-1},x', x_{j+1},\dots, x_N), \\
\hat{\bf{x}}_{j,k} & = (x_1,\dots,x_{j-1},x, x_{j+1},\dots,x_{k-1}
,x',x_{k+1},\ldots, x_N), \\
d\hat{\bf{x}}_{j,k} &= dx_1\dots dx_{j-1} dx_{j+1}\dots
dx_{k-1} dx_{k+1} \dots dx_N.
\end{align*}
Then $\gamma_1$ and $\rho_2$ are defined by
\begin{align}
\gamma_1(x,x') &= \sum_{j=1}^N \int_{{\mathbb R}^{3N-3}} \psi(\hat{\bf{x}}_j) 
\overline{\psi(\hat{\bf{x}}'_j)} \,d\hat{\bf{x}}_{j}, \\
\rho_2(x,x') & = \sum_{1\leq j \neq k \leq N} \int_{{\mathbb R}^{3N-6}}
|\psi(\hat{\bf{x}}_{j,k})|^2\,d\hat{\bf{x}}_{j,k}.
\end{align}
In order to describe the regularity of $\gamma_1$ and $\rho_2$ we 
introduce $D=\{(x,x) \in {\mathbb R}^6\}$ and 
$$
S=\left(\{R_1,\ldots,R_L\}\times {\mathbb R}^3\right) \cup
        \left({\mathbb R}^3 \times\{R_1,\ldots,R_L\}\right) \subset {\mathbb R}^6.
$$
Our method implies that $\gamma_1$ is real analytic on 
${\mathbb R}^6 \setminus S$ and that $\rho_2$ is real analytic on 
${\mathbb R}^6 \setminus (D \cup S)$. 
\end{remark}

\begin{remark}
In the case of an atom, consider the density $\rho$ in polar coordinates ($x=r\omega, \,
r=|x|, \omega=x/|x|\in\mathbb S^2$), and define 
$\tilde \rho(r)=\int _{\mathbb S^2}\rho(r\omega)d\omega$. 
An important question is
for which $k\in\mathbb N$, 
\begin{equation}
\label{eq:deriv_0}
\bigg(\frac{d^k\tilde \rho}{dr^k}\bigg)(0)
\end{equation} 
exists. An even more demanding question is whether $\tilde\rho(r)$
is real analytic for $r\ge 0$, i.e. whether $\tilde{\rho}$ can be continued analytically beyond $0$. 
The analysis of such questions show the intimate relation
between the problems i) and ii). 
In \cite{HHOS:2001} it was shown that the derivatives in \eqref{eq:deriv_0} 
exist for $k\leq 2$. But the 
general problem remains open.  
\end{remark}

\begin{remark}[Generalisations]
\label{rem:generalise}
As will be seen from the proof, Theorem \ref{thm:density_analytic} can easily be generalised to many other potentials. 
We do not use any of the special properties of Coulomb potentials, such as symmetry, homogeneity, etc.
To be precise, let
$$
V({\bf x}) = \sum_{j=1}^N V_j(x_j) + \sum_{j,k=1,j\neq k}^N W_{j,k}(x_j-x_k),
$$
satisfy the following conditions:
\begin{itemize}
\item[(1)] There exists $C >0$ such that for all $u \in
W^{1,2}({\mathbb R}^{3N})$, 
\begin{equation}
\label{eq:Hardy_generalized}
\| V u \|_{L^{2}({\mathbb R}^{3N})} \leq C \| u
\|_{W^{1,2}({\mathbb R}^{3N})}.
\end{equation}
\item[(2)] There exists a constant $L > 0$ (depending on
$\varepsilon$) such that for all $\alpha \in {\mathbb N}^{3}$, we have
\begin{equation}
\label{eq:analyticity_V_generalized}
\sum_{j=1}^N \| \partial^{\alpha} V_j \|
+
\sum_{j,k=1, j\neq k}^N \| \partial^{\alpha} W_{j,k} \|
\leq L^{|\alpha|+1} |\alpha|!,
\end{equation}
where the norms in \eqref{eq:analyticity_V_generalized} are in 
$L^{\infty}\left(\left\{x \in {\mathbb R}^3 \,\big{|}\,|x|>\varepsilon \right\}
\right)$.
\end{itemize}
The first condition, \eqref{eq:Hardy_generalized}, is a kind of relative boundedness assumption. The second condition, \eqref{eq:analyticity_V_generalized}, means that $V$ is real analytic away from $\Sigma$ (with a uniformity at infinity).
Theorem \ref{thm:density_analytic} remains true for any $V$ satisfying these two assumptions . For instance, replacing one or more of the Coulomb potentials in $V$ by the Yukawa potential, $e^{-\alpha |x|}/|x|$ (with $\alpha >0$), we still get Theorem \ref{thm:density_analytic}. But here we concentrate on the physically important case of Coulomb potentials and do not strive for generality.
\end{remark}

{\it Organisation of the paper:}\\
In section \ref{section3} we present in Lemma \ref{lem:diff_parallel}
a result concerning `partial analyticity' of an eigenfunction $\psi$ of $H$
in the following sense: Upper bounds to the  $L^2$-norms
of certain directional derivatives of $\psi$ of arbitrary
order $|\alpha|$ are given. They show the right behaviour in $|\alpha|$
needed later on for the proof of the analyticity of $\rho$ away from 
the origin. (The proof of Lemma \ref{lem:diff_parallel} is given in 
Appendix \ref{analyticity_proof}.)
We note that this kind of directional derivatives correspond,
roughly speaking, to `taking derivatives along singularities
of the potential', see Lemma \ref{lem:Properties_of_V} and its proof, and compare also with
\cite{FHHO} and \cite{FHHO1:2002}.
Corollary \ref{cor:diff_parallel} is an immediate consequence of Lemma \ref{lem:diff_parallel} and essential
for the further steps in the proof of Theorem \ref{thm:density_analytic}.

In section \ref{section4} we state and prove Proposition
\ref{prop:analytic_estimate}
which gives us 
the necessary control on $|(\partial^\alpha\rho)(x)|$ for $|x|\ge 
\epsilon>0$  and arbitrary $\alpha$. Therefrom the analyticity of $\rho$
follows immediately.  The key point of the proof of Proposition \ref{prop:analytic_estimate} is
Lemma \ref{lem:differentiation_under}. For its proof we use a suitable partition of unity of
$\mathbb R^{3N}$ and then proceed by a similar construction as in \cite{FHHO}
 which together with Corollary \ref{cor:diff_parallel} implies 
Lemma \ref{lem:differentiation_under}, in particular 
\eqref{eq:diff_finale}. 
Once Lemma \ref{lem:differentiation_under} is proved, Proposition
\ref{prop:analytic_estimate} follows by easy arguments.

\section{Basic facts and notation}
Remember that for multiindices $\alpha = (a_1,\ldots,a_{3M}) \in
{\mathbb N}^{3M}$,
$$
|\alpha | = \sum_{j=1}^{3M} a_j.
$$
Furthermore, we have the usual ordering on multiindices: For $\alpha =
(a_1,\ldots,a_{3M})$, $\beta=(b_1,\ldots,b_{3M})$ we write $\alpha
\leq \beta$ iff $a_j \leq b_j$ for all $j$.

We will need one simple and standard combinatorical fact. We recall it here for the reader's convenience.
\begin{prop}
\label{prop:combinatorics}
  Let $\alpha \in {\mathbb N}^{3M}$ be a multiindex. Then
$$
  \sum_{\beta \leq \alpha, |\beta| = b} \binom{\alpha}{\beta} =
  \binom{|\alpha|}{b}.
$$
\end{prop}
Proposition \ref{prop:combinatorics} will be used as follows. Use
Leibniz' rule to calculate
$$
\partial^{\alpha} (f g)= \sum_{\beta \leq \alpha} \binom{\alpha}{\beta}
(\partial^{\beta}f) (\partial^{\alpha-\beta}g).
$$
Then the number of terms where exactly $b$ differentiations fall on
$f$ is given by $\binom{|\alpha|}{b}$.

In the following we shall work with certain directional derivatives. Let
$e_s$ for $s \in \{1,2,3\}$ denote the standard basis for ${\mathbb R}^3$. Let $P$ be a (non-empty) subset of $\{1,\ldots,N\}$. 
We define the coordinate $x_P$ by
$$
x_P = \frac{1}{\sqrt{|P|}} \sum_{j\in P} x_j.
$$
We will now define
$\partial_{x_P}^{e_s} f$ for a function $f \in C^1({\mathbb
  R}^{3N})$. 
For the given $P$ and $s$ let ${\mathbf v}=(v_1,\ldots,v_N) \in
{\mathbb R}^{3N}$ with $v_j = 0$ for $j \notin P$, and $v_j = e_s/\sqrt{|P|}$ for $j
\in P$.
Then we define
$$
\partial_{x_P}^{e_s} f(x) = \nabla f \cdot {\mathbf v}.
$$
The definition of $\partial_{x_P}^{\alpha}$ then follows
by iteration
for any $\alpha \in {\mathbb N}^3$. 
One can clearly reformulate this definition in terms of Fourier
transforms (multiplication by $\xi_{P}^{e_s}$ for suitably defined
$\xi_{P}$ in Fourier space). In the previous paper \cite{FHHO} we used a coordinate transformation to describe these derivatives.

\section{Partial analyticity of atomic eigenfunctions}
\label{section3}
We will need a result on partial analyticity of the
eigenfunctions of $H$.

\begin{lemma}
\label{lem:diff_parallel}
Let $\psi \in L^2({\mathbb R}^{3N})$ be an eigenfunction of
$H$.
Let the index sets 
$P_1,\ldots,P_M \subset \{1,\ldots,N\}$ satisfy for all
$s
\in
   \{1,\ldots,M\}$ that $P_s \neq \emptyset$.
   Define for each $s$, $Q_s = \{1,\ldots,N\} \setminus P_s$.
Define also, for $\varepsilon > 0$,
\begin{multline}
   \label{eq:def_U_P_s}
   U_{P_s}(\varepsilon) = \Big\{
(x_1,\ldots,x_N) \in{\mathbb R}^{3N}\,\Big|\, |x_j| > \varepsilon \mbox{
   for } j \in P_s,  \\  |x_j-x_k| > \varepsilon \mbox{ for } j
\in P_s, k \in Q_s \Big\}.
\end{multline}
Denote
\begin{align}
\label{eq:def_U_P}  U_{P_1,\ldots,P_M}(\varepsilon) = \cap_{s=1}^M 
U_{P_s}(\varepsilon).
\end{align}
Then there exist $C,L$ (depending on $\varepsilon$) such that
for all multiindices,
$\alpha = (\alpha_1,\ldots,\alpha_M) \in{\mathbb N}^{3M}$, we
have
$$
\|\partial^{\alpha_1}_{x_{P_1}}
\cdots\partial^{\alpha_M}_{x_{P_M}}\psi
\|_{L^2(U_{P_1,\ldots,P_M}(\varepsilon))}\leq
C L^{|\alpha|}(|\alpha|+1)^{|\alpha|}.
$$
\end{lemma}

The proof of Lemma \ref{lem:diff_parallel} is similar to
the standard proof that solutions to elliptic equations with
analytic coefficients are analytic 
(see \cite[Section 7.5, pp. 177-180]{H0}) and will be given in
Appendix \ref{analyticity_proof}.

Let us introduce the following
practical notation. For a multiindex
$\alpha = (\alpha_1,\ldots,\alpha_M)\in {\mathbb N}^{3M}$ and given
$P_1, \ldots, P_M$ as in Lemma \ref{lem:diff_parallel}, we
define
$\partial_{x_{{\bf P}}}^{\alpha}$ and $U_{\mathbf P}(\varepsilon)$ by
\begin{align}
\label{eq:abusive_notation}
  \partial_{x_{{\mathbf P}}}^{\alpha} =
\partial^{\alpha_1}_{x_{P_1}}
\cdots\partial^{\alpha_M}_{x_{P_M}},
\,\,\,\,\,\,\,\,\,\,\,\,
U_{\mathbf P}(\varepsilon)=U_{P_1,\ldots,P_M}(\varepsilon).
\end{align}

We will need the result of Lemma \ref{lem:diff_parallel} in a slightly
different form for the proof of Theorem
\ref{thm:density_analytic}. For later convenience, we state and prove
this reformulation here.

\begin{cor}
  \label{cor:diff_parallel}
  Let the notation and assumptions be as in Lemma
  \ref{lem:diff_parallel} (using \eqref{eq:abusive_notation}). Then
  there exist constants $C_1, L_1$ such that
$$
\int_{U_{\mathbf P}(\varepsilon)} \left| \partial_{x_{{\mathbf
        P}}}^{\alpha} |\psi|^2({\mathbf x}) \right| \,d{\mathbf x}
\leq
C_1 L_1^{|\alpha|} (|\alpha|+1)^{|\alpha|}.
$$
\end{cor}

\begin{proof}
By Leibniz' rule, we have
\begin{align}
  \label{eq:Leibniz}
\partial_{x_{{\mathbf P}}}^{\alpha} |\psi|^2 
=
\sum_{\beta \leq \alpha} \binom{\alpha}{\beta} 
\overline{\partial_{x_{{\mathbf P}}}^{\beta} \psi} \,
\partial_{x_{{\mathbf P}}}^{\alpha-\beta} \psi.
\end{align}
Applying Cauchy-Schwarz and Lemma \ref{lem:diff_parallel} to both 
$\overline{\partial_{x_{{\mathbf
      P}}}^{\beta} \psi}$ and $\partial_{x_{{\mathbf
      P}}}^{\alpha-\beta} \psi$ in \eqref{eq:Leibniz}, 
we find using Proposition \ref{prop:combinatorics}
for the equality below,
\begin{align*}
  \int_{U_{\mathbf P}(\varepsilon)} &\left| \partial_{x_{{\mathbf
        P}}}^{\alpha} |\psi|^2({\mathbf x}) \right| \,d{\mathbf x} \\
&\leq
C^2 L^{|\alpha|} \sum_{\beta \leq \alpha} \binom{\alpha}{\beta}
(|\beta|+1)^{|\beta|} (|\alpha|-|\beta|+1)^{|\alpha|-|\beta|} \\
  &= C^2 L^{|\alpha|} \sum_{b=0}^{|\alpha|} \binom{|\alpha|}{b}
  (b+1)^b (|\alpha|-b+1)^{|\alpha|-b} \\
  & \leq
  C^2 (2 L)^{|\alpha|} (|\alpha| + 1)^{|\alpha|}.
\end{align*}
Thus, Corollary \ref{cor:diff_parallel} holds with $C_1=C^2$, $L_1=2L$.
\end{proof}

\section{Differentiating the density}
\label{section4}
Fix an arbitrary $\varepsilon > 0$. We will always
study $\rho(x_1)$ in the region $\{|x_1| > \varepsilon\}$. We
will prove the following estimate:

\begin{prop}
\label{prop:analytic_estimate}
Let $\varepsilon >0$ be given. Then there exist constants $C, L > 0$, such that for all $|x| > \varepsilon$ and all $\alpha \in {\mathbb N}^3$, $\rho$ satisfies
\begin{equation}
\label{eq:analytic_constants}
\left| \partial^{\alpha} \rho(x) \right| \leq C
L^{|\alpha|}(|\alpha|+1)^{|\alpha|}.
\end{equation}
\end{prop}

\begin{remark}
It is clear that Proposition \ref{prop:analytic_estimate} implies Theorem \ref{thm:density_analytic}.
\end{remark}

\begin{proof}[Proof of Proposition \ref{prop:analytic_estimate}]
Choose $\chi_1, \chi_2 \in C^{\infty}({\mathbb R}^{3})$,
satisfying
$$
\chi_1+ \chi_2 = 1, \,\,\,\,\,\,
\chi_1 \equiv 1 \text{ on } B(0,\varepsilon/(4N)),\,\,\,\,\,\,
\supp \chi_1 \subset
B(0,\varepsilon/(2N)),
$$
and let further $\chi_1, \chi_2$ be radially symmetric functions.
Using this partition of unity and the notation
$$
{\mathcal M} = \left\{(j,k) \in \{1,\ldots,N\}^2 \,\Big{|}\, j<k\right\},
$$
we can write
\begin{eqnarray}
\label{eq:phi_I}
\rho(x_1) &=& \int |\psi({\bf x})|^2
\prod_{j<k} \left(\chi_1(x_j-x_k)+ \chi_2(x_j-x_k)\right)
dx_2\cdots
dx_N \nonumber \\
&=&
\sum_{I \subset {\mathcal M} } \int |\psi|^2({\bf
x}) \phi_I({\bf x})
\,dx_2\cdots
dx_N \nonumber \\
&\equiv& \sum_{I \subset {\mathcal M} } \rho_I(x_1). 
\end{eqnarray}
Equation \eqref{eq:phi_I} defines $\phi_I$ as
\begin{equation}
\label{eq:phi_I_def}
\phi_I = \Big\{\prod_{(j,k) \in I} \chi_1(x_j-x_k)\Big\}
\Big\{\prod_{(j,k) \in {\mathcal M} \setminus I} \chi_2(x_j-x_k)\Big\}.
\end{equation}
We will prove that $\rho_I(x_1)$ satisfies an estimate like
\eqref{eq:analytic_constants} on $\{|x_1| > \varepsilon\}$ for
all $I \subset {\mathcal M}$, namely
\begin{equation}
\label{eq:analytic_constants_terms}
|\partial^{\alpha} \rho_{I}| \leq C
L^{|\alpha|}(|\alpha|+1)^{|\alpha|}.
\end{equation}

The estimate \eqref{eq:analytic_constants} follows from
\eqref{eq:analytic_constants_terms} (with a different $C$) since the 
sum in \eqref{eq:phi_I} is finite. 

The estimate \eqref{eq:analytic_constants_terms} is a consequence of 
\eqref{eq:diff_finale} (with $\phi = \phi_I$) in Lemma \ref{lem:differentiation_under} below, using a Sobolev imbedding theorem. Since we have not found an ideal reference we include the following easy argument:

Let $v \in C^{\infty}({\mathbb R}^3)$, $v(x) =1 $ for $|x| \geq \varepsilon$, $v(x) = 0$ for $|x|\leq \varepsilon/2$. Let furthermore ${\mathcal F}$ denote the Fourier transformation. We get for $\alpha \in {\mathbb N}^3$, $|x| \geq \epsilon$,
\begin{align*}
\partial^{\alpha} \rho(x) &= (v \partial^{\alpha} \rho)(x) \\
        &= c \int e^{ixp} (1+p^2)^{-2} 
        {\mathcal F}\left( (1-\Delta)^2 \left(v \partial^{\alpha} \rho\right) \right) \, dp.
\end{align*}
Therefore,
\begin{align}
\label{eq:fourier}
\left| \partial^{\alpha} \rho(x) \right| 
\leq
c \|(1+p^2)^{-2}\|_{L^1({\mathbb R}^3)} 
\| (1-\Delta)^2 (v \partial^{\alpha} \rho) \|_{L^1({\mathbb R}^3)}.
\end{align}
We can estimate 
$\| (1-\Delta)^2 (v \partial^{\alpha} \rho) \|_{L^1({\mathbb R}^3)}$ using 
\eqref{eq:diff_finale} by,
\begin{align}
\label{eq:fourier_estimate}
\| (1-\Delta)^2 (v \partial^{\alpha} \rho) \|_{L^1({\mathbb R}^3)}
&\leq c_1 L_1^{|\alpha|+4} (|\alpha|+4 +1)^{|\alpha|+4 } \nonumber \\
&\leq c_2 L_2^{|\alpha|} (|\alpha|+1)^{|\alpha| },
\end{align}
for some constants $c_1, L_1, c_2, L_2$.
Combining \eqref{eq:fourier} and \eqref{eq:fourier_estimate} yields
\eqref{eq:analytic_constants_terms}.
Proving Lemma \ref{lem:differentiation_under} therefore finishes the proof of Proposition \ref{prop:analytic_estimate}.
\end{proof}

\begin{lemma}
\label{lem:differentiation_under}
Let $\varepsilon>0$ be given and let
\begin{align}
  \label{eq:phi}
  \phi = \prod_{1\leq j<k\leq N} f_{j,k}(x_j - x_k),
\end{align}
where each $f_{j,k}$ is one of the functions
$\chi_1,\chi_2, \partial^{e_s}\chi_2$, with $e_s \in
{\mathbb N}^3$, $|e_s|=1$.
\begin{itemize}
\item[{\bf i)}]
Let $P_1,\ldots,P_M$ be subsets of $\{1,\ldots,N\}$
satisfying that $1 \in P_j$ for $j=1,\ldots, M$
and
\begin{align}
  \label{eq:support_M}
(\supp \phi) \cap \{ |x_1| > \varepsilon \} \subset
\cap_{j=1}^M U_{P_j}(\varepsilon/(4N)).
\end{align}
Then there exist constants $C,L >0$ (depending on $\varepsilon$) 
such that for
all multiindices, $\alpha = (\alpha_0,\alpha_1,\ldots,\alpha_M) \in
{\mathbb N}^{3M+3}$, we have
\begin{multline}
\label{eq:foerste_led}
\left\| \partial_{x_1}^{\alpha_0} \int
(\partial_{x_{P_1}}^{\alpha_1}
\cdots \partial_{x_{P_M}}^{\alpha_M} |\psi|^2)({\bf x})
\phi({\bf x}) \,dx_2 \cdots dx_N \right\|_{L^1(\{|x_1|>\varepsilon\})} \\
\leq
C L^{|\alpha|} (|\alpha|+1)^{|\alpha|}.
\end{multline}
\item[{\bf ii)}]
There exist constants $C,L >0$ (depending on $\varepsilon$)
such that for all $\alpha \in {\mathbb N}^{3}$ we have
\begin{equation}
\label{eq:diff_finale}
\left\|
\partial^{\alpha}_{x_1}\int |\psi|^2({\bf
x}) \phi({\bf x}) \,dx_2\cdots dx_N \right\|_{L^1(\{|x_1|>\varepsilon\})}
\leq C L^{|\alpha|} (|\alpha|+1)^{|\alpha|}.
\end{equation}
\end{itemize}
\end{lemma}

\begin{proof}
To a function $\phi$ as given in \eqref{eq:phi} we
will associate $P = P(\phi) \subset \{1,\ldots,N\}$
satisfying $1\in P$ and such that
\begin{align}
  \label{eq:support}
\left( \supp \phi \right) \cap \{|x_1| > \varepsilon\}
\subset U_{P(\phi)}(\varepsilon/(4N)).
\end{align}
We will now 
describe the map $\phi \mapsto P(\phi)$.
We note that the following construction is similar to the one from \cite{FHHO}.
Define $I = I(\phi) \subset {\mathcal M}$ by
$$
(j,k) \in I(\phi) \text{ if and only if }
  f_{j,k} \in \{\chi_1,
\partial^{e_1}\chi_2,\partial^{e_2}\chi_2,
\partial^{e_3}\chi_2\}.
$$
In other words, $(j,k) \in I(\phi)$ means precisely that $f_{j,k} \neq
\chi_2$. 
The set $I(\phi)$ generates an equivalence relation on $\{1,\ldots,N\}^2$ and we define $P(\phi)$ to be the equivalence class of $1$.
Less abstractly, this means that
\begin{itemize}
\item $1 \in P(\phi)$.
\item For $j \geq 2$ we have
$j \in P(\phi)$ iff there exists $\{j_1,\ldots,j_s\}
\subseteq \{1,\ldots,N\}$, $s \leq N$, satisfying
   \begin{itemize}
   \item[] $(1,j_1) \in I(\phi)$,
   \item[] $(j_t,j_{t+1}) \in I(\phi)$ or $(j_{t+1},j_{t}) \in
    I(\phi)$ for $1\leq t < s$,
\end{itemize}
and
\begin{itemize}
        \item[] $(j_s,j) \in I(\phi)$ or $(j,j_{s}) \in
    I(\phi)$.
   \end{itemize}
\end{itemize}
Notice that, since $\chi_1 + \chi_2 = 1$, $\supp
\partial^{e_j}\chi_2 \subset \supp \chi_1$, $j=1,2,3$.
Therefore we get \eqref{eq:support} by the same elementary geometrical
considerations (the triangle inequality) as in \cite{FHHO}.

In the proof we shall use $P = P(\phi)$ in order to replace the
derivative $\partial_{x_1}^{\alpha_0}$ outside the integral in the
left hand side of \eqref{eq:foerste_led} by the derivative
$\partial_{x_P}^{\alpha_0}$ inside the integral. That will enable us 
to apply Corollary \ref{cor:diff_parallel}.

Let $P=P(\phi)$ according to our construction. We will prove the lemma
recursively in $|P|$. In the proof below we will freely interchange 
the order of differentiation (in the distributional sense) 
and integration. This is permitted, due to Corollary 
\ref{cor:diff_parallel}, which ensures that the derivatives of  
the functions in question belong to 
$L^1\left(\{|x_1|>\epsilon\} \times{\mathbb R}^{3N-3}\right)$. 
We will only prove part ${\bf i)}$ of Lemma \ref{lem:differentiation_under}. 
The changes necessary for the case ${\bf ii)}$ are obvious and therefore 
omitted. 

{\bf Step 1, $|P|=N$.} In the case where $P=\{1,\ldots,N\}$
we make the change of variables $y_j = x_j -x_1$ for
$j=2,\ldots,N$. Then we get $x_j-x_k = y_j -y_k$ for $j,k
\neq 1$.
The point is that $\phi$ only depends on the differences
$x_j-x_k$, and therefore, after the change of variables, the
only dependence on $x_1$ will be in $|\psi|^2$, where we can
apply Corollary \ref{cor:diff_parallel}. Let us carry this out.

Denote
${\bf y} = (y_2,\dots,y_N)$. Then we see that after change
of variables we have
$$
\phi({\bf x}) = \tilde{\phi}({\bf y}),
$$
for some function $\tilde{\phi}$. 
Explicitly, we see from \eqref{eq:phi} that
$$
\tilde{\phi}(y_2,\ldots,y_N) = \prod_{1<j<k} f_{j,k}(y_j-y_k) 
\prod_{s=2}^N f_{1,s}(-y_s).
$$
Therefore,
\begin{eqnarray}
\label{eq:induktion_N}
&&\int
(\partial_{x_{P_1}}^{\alpha_1}
\cdots \partial_{x_{P_M}}^{\alpha_M} |\psi|^2)({\bf x})
\phi({\bf x}) \,dx_2 \cdots dx_N \\
&=&
\int
(\partial_{x_{P_1}}^{\alpha_1}
\cdots \partial_{x_{P_M}}^{\alpha_M} |\psi|^2)(x_1,
x_1+y_2,\ldots, x_1+y_N)
\tilde{\phi}({\bf y}) \,d{\bf y}.\nonumber
\end{eqnarray}
 From \eqref{eq:induktion_N} we get by differentiation under
the integral sign and change of coordinates back to ${\mathbf x}$:
\begin{eqnarray}
\label{eq:induktion_N_diff_under}
&&\partial_{x_1}^{\alpha_0}\int
(\partial_{x_{P_1}}^{\alpha_1}
\cdots \partial_{x_{P_M}}^{\alpha_M} |\psi|^2)({\bf x})
\phi({\bf x}) \,dx_2 \cdots dx_N \\
&=&
\int
(\partial_{x_P}^{\alpha_0}\partial_{x_{P_1}}^{\alpha_1}
\cdots \partial_{x_{P_M}}^{\alpha_M} |\psi|^2)(x_1,
x_1+y_2,\ldots, x_1+y_N)
\tilde{\phi}({\bf y}) \,d{\bf y}\nonumber \\
&=&
\int
(\partial_{x_P}^{\alpha_0}\partial_{x_{P_1}}^{\alpha_1}
\cdots \partial_{x_{P_M}}^{\alpha_M} |\psi|^2)({\mathbf x})
\phi({\bf x}) \,dx_2 \cdots dx_N.\nonumber
\end{eqnarray}
Notice the support conditions \eqref{eq:support}, \eqref{eq:support_M}.
We can now apply Corollary \ref{cor:diff_parallel} to get
\eqref{eq:foerste_led} 
in the case $|P|=N$.

{\bf Step 2, $|P|<N$.} 
Suppose that Lemma \ref{lem:differentiation_under} holds under the additional 
assumption $|P|>K$ for some $0 \leq K < N$. We will prove the statement for  
$|P|=K$.

Define $Q= \{1,\ldots,N\}\setminus P$. Since
$|P| < N$, $Q \neq \emptyset$. 
Note that if $j \in P$, $k \in Q$ then, by definition of $I(\phi)$ and
$P = P(\phi)$, 
we have $(j,k) \notin I(\phi)$ and $(k,j) \notin I(\phi)$. Therefore, if $j<k$ we have $f_{j,k} = \chi_2$ and if $k < j$ we have $f_{k,j} = \chi_2$.
So
$\phi$ contains the factor (remember that $\chi_2$ 
is rotationally symmetric, in particular even)
$$
\phi_{P,Q} = \prod_{j \in P, k \in Q} \chi_2(x_j-x_k),
$$
and can be written as
$$
\phi = \phi_{P} \cdot \phi_Q \cdot \phi_{P,Q},
$$
where
\begin{eqnarray*}
   \phi_{P} = \prod_{j,k \in P, j<k} f_{j,k}(x_j-x_k), 
\,\,\,\,\,\,\,\,\,\,\,\,
   \phi_{Q} = \prod_{j,k \in Q, j<k} f_{j,k}(x_j-x_k).
\end{eqnarray*}
We do the following change of variables for $j \geq 2$:
$$
y_j = \begin{cases} x_j-x_1 & \text{ for } j \in P \setminus \{1\}, \\
                     x_j & \text{ for } j \in Q. \end{cases}
$$
For convenience of notation we define $y_1=0$.
We clearly get for $j,k \in P$ or $j,k \in Q$,
that
$x_j-x_k = y_j-y_k$ (also when either $j=1$ or $k=1$---remember that $1 \in P$). 
Thus, as in the case $|P|=N$, we can
write $\phi_{P}({\bf x}) = \tilde{\phi}_P({\bf y})$, and
$\phi_{Q}({\bf x}) = \tilde{\phi}_Q({\bf y})$.

Write ${\bf
z}=(z_1,z_2,\ldots,z_N)
\in {\mathbb R}^{3N}$ with $z_j =
\begin{cases} x_1 & j=1, \\
               x_1+y_j & j \in P \setminus\{1\}, \\
               x_j & j \in Q. \end{cases}$
Then
\begin{align}
\label{eq:changed_coord}
   &\int
(\partial_{x_{P_1}}^{\alpha_1}
\cdots \partial_{x_{P_M}}^{\alpha_M} |\psi|^2)({\bf x})
\phi({\bf x}) \,dx_2 \cdots dx_N \\
=&
\int
(\partial_{x_{P_1}}^{\alpha_1}
\cdots \partial_{x_{P_M}}^{\alpha_M} |\psi|^2)({\bf z})
\tilde{\phi}_P({\bf y})
\left( \prod_{j\in P, k\in Q}\chi_2 (x_1+y_j-y_k)\right)
\tilde{\phi}_Q({\bf y}) \,d{\bf y}. \nonumber 
\end{align}
Differentiation under the integral sign yields
\begin{align}
\label{eq:nabla}
   &\nabla_{x_1} \int
(\partial_{x_{P_1}}^{\alpha_1}
\cdots \partial_{x_{P_M}}^{\alpha_M} |\psi|^2)({\bf x})
\phi({\bf x}) \,dx_2 \cdots dx_N  \\
=&
\int
(\nabla_{x_P} \partial_{x_{P_1}}^{\alpha_1}
\cdots \partial_{x_{P_M}}^{\alpha_M} |\psi|^2)({\bf z})
\tilde{\phi}_P({\bf y})\,
\big\{ \!\!\!\!\!\!
\prod_{j\in P, k\in Q}\chi_2(x_1+y_j-y_k)
\big\}
\tilde{\phi}_Q({\bf y}) \,d{\bf y}\nonumber \\ &
+ \sum_{j \in P, k\in Q}  \int
(\partial_{x_{P_1}}^{\alpha_1}
\cdots \partial_{x_{P_M}}^{\alpha_M} |\psi|^2)({\bf z})
\tilde{\phi}_P({\bf y})\tilde{\phi}_Q({\bf y})\nonumber \\
&\times
\Big( \prod_{j'\in P, k'\in Q,(j',k')\neq(j,k)}
\chi_2(x_1+y_{j'}-y_{k'})\Big)
(\nabla\chi_2)(x_1+y_j-y_k)\,d{\bf y}. \nonumber
\end{align}
Let us explain roughly how we will proceed for higher derivatives 
with respect to $x_1$. 
For each consecutive differentiation we will get terms as in \eqref{eq:nabla}.
The term where all differentiations fall on $\psi$ can be differentiated again in
a manner similar to \eqref{eq:nabla}. If one differentiation falls on $\chi_2$ 
we stop differentiating that term under the integral sign---leaving the rest of 
the differentiations outside the integral. The result of this procedure is 
\eqref{eq:induction}, the notation of which we will define below. The important point 
is that when all derivatives fall on $\psi$, we can apply Corollary
\ref{cor:diff_parallel} to obtain our conclusion. On the other hand a
differentiation of $\chi_2$ will lead to a situation with a larger
$|P|$---so these terms can be handled by the induction 
hypothesis.

Let $\eta_s \in {\mathbb N}^3$, $|\eta_s|=1$, $s=1,2,\ldots,S$.
Define for $t\in \{1,\ldots ,S\}$
\begin{align*}
  A_0 &= 0, &
A_t &= \sum_{s\leq t} \eta_s, &
B_t &= \sum_{s=t+1}^S \eta_s, &
B_S &= 0.
\end{align*}
Notice that the definition of $B_t$ depends on $S$, i.e. $B_t = B_t(S)$.
We get the following formula \eqref{eq:induction} from \eqref{eq:nabla}, 
using the procedure described above, by induction with respect to $S$. 
For $S=1$ the equation \eqref{eq:induction} reduces to
\eqref{eq:nabla}.
\begin{align}
\label{eq:induction}
   \partial_{x_1}^{A_S} \Big\{
\int
(\partial_{x_{P_1}}^{\alpha_1}
&\cdots \partial_{x_{P_M}}^{\alpha_M} |\psi|^2)({\bf x})
\phi({\bf x}) \,dx_2 \cdots dx_N \Big\}  \\
=
\int
(\partial_{x_P}^{A_S}\partial_{x_{P_1}}^{\alpha_1}
&\cdots \partial_{x_{P_M}}^{\alpha_M} |\psi|^2)({\bf x})
\phi({\bf x}) \,dx_2 \cdots dx_N  \nonumber \\
+
\sum_{t=1}^S \partial_{x_1}^{B_t}
\Big\{ &
\sum_{j \in P, k\in Q}\int
(\partial_{x_P}^{A_{t-1}}\partial_{x_{P_1}}^{\alpha_1}
\cdots \partial_{x_{P_M}}^{\alpha_M} |\psi|^2)({\bf z})
\tilde{\phi}_P({\bf y}) \tilde{\phi}_Q({\bf y}) \nonumber \\
&\times
\Big( \prod_{j'\in P, k'\in
    Q,(j',k')\neq(j,k)}\chi_2(x_1+y_{j'}-y_{k'})\Big) \nonumber \\
&
\,\,\,\,\,\,\,\,\,\,\,\,\,\,\,\,\,\,\,
\,\,\,\,\,\,\,\,\,\,\,\,\,\,\,\,\,\,\,
\times
(\partial^{\eta_t}\chi_2)(x_1+y_j-y_k)\,dy_2\cdots dy_N
\Big\}.\nonumber
\end{align}
We will use \eqref{eq:induction} with $S=|\alpha_0|$, $A_S = \alpha_0$.
Consider the function
$$
\phi_{j,k} = \phi_P \cdot \phi_Q \cdot
(\partial^{\eta_t}\chi_2)(x_j-x_k)\prod_{j'\in P, k'\in
   Q,(j',k')\neq(j,k)}\chi_2(x_{j'}-x_{k'}).
$$
By construction we have $|P(\phi_{j,k})| > |P(\phi)|$.
Therefore, we get by the induction hypothesis on $|P|$ that
\begin{align*}
&\left\|
\partial_{x_1}^{B_t} \int
(\partial_{x_P}^{A_{t-1}}\partial_{x_{P_1}}^{\alpha_1}
\cdots \partial_{x_{P_M}}^{\alpha_M} |\psi|^2)({\bf x})
\phi_{j,k}({\bf x}) \,dx_2 \cdots dx_N\right\|_{L^1(\{|x_1|>\varepsilon\})} \\
& \leq 
C L^p (p+1)^p,
\end{align*}
where $p = |B_t| + |A_{t-1}| + |\alpha_1| +\ldots+ |\alpha_M| =|\alpha|-1$.
Furthermore, using Corollary \ref{cor:diff_parallel} on the first term on the
right hand side in \eqref{eq:induction}, we obtain
$$
\left\| \int
(\partial_{x_P}^{A_S}\partial_{x_{P_1}}^{\alpha_1}
\cdots \partial_{x_{P_M}}^{\alpha_M} |\psi|^2)({\bf x})
\phi({\bf x}) \,dx_2 \cdots dx_N \right\|_{L^1(\{|x_1|>\varepsilon\})}
\leq
C L^p (p+1)^p,
$$
with $p = |A_S| + |\alpha_1| +\ldots+ |\alpha_M|= |\alpha|$. 

Thus the desired estimate holds for the individual terms on the right hand side 
in \eqref{eq:induction}. Since the number of terms is bounded by $c |\alpha|$ 
this finishes the
proof of \eqref{eq:foerste_led}.
\end{proof}

\appendix
\section{Proof of Lemma \ref{lem:diff_parallel}}
\label{analyticity_proof}

In this appendix we will prove Lemma
\ref{lem:diff_parallel}. 
For convenience define
$H_E = H-E$, with $E$ being the eigenvalue corresponding to the 
eigenfunction $\psi$, i.e. $\psi$ satisfies 
$H_E \psi =0$.
Recall the notations given in \eqref{eq:abusive_notation}.

Let us start the proof by stating a well known result
explicitly. Since the domain of
$H_E$ is known to be $W^{2,2}({\mathbb R}^{3N})$, we get

\begin{lemma}
\label{lem:obvious_domain}
Let $v \in  W^{1,2}({\mathbb R}^{3N})$. Then 
$v \in W^{2,2}({\mathbb R}^{3N})$ if and only if
$H_E v \in L^2({\mathbb R}^{3N})$.
Furthermore, there exists a constant $K_0 >0$ such that for all 
$v \in W^{2,2}({\mathbb R}^{3N})$
$$
\| v \|_{W^{2,2}({\mathbb R}^{3N})} \leq
K_0\left( \|H_E v \|_{L^2({\mathbb R}^{3N})} +
\|v \|_{L^2({\mathbb R}^{3N})} \right).
$$
\end{lemma}
This follows from the fact that $V$ is infinitesimally small (in the operator sense) with respect to $-\Delta$.

We now state and prove an {\it a priori} estimate.
\begin{lemma}[A priori estimate]
\label{lem:a_priori}
There exists a constant $C_0$ such that for all $\eta,\eta_1
\in (0,1)$, all $\alpha_0 \in {\mathbb N}^{3N}$, 
$\alpha' \in {\mathbb N}^{3M}$ with
$|\alpha_0|+ |\alpha'| \leq 2$ and all 
$v \in W^{2,2}(U_{{\bf P}}(\varepsilon/2+\eta_1))$ we
have the estimate
\begin{align}
\label{eq:dumt_nummer}
\eta^{|\alpha_0|+|\alpha'|}& \| \partial^{\alpha_0}\partial^{\alpha'}_{x_{{\bf P}}}
v\|_{L^2(U_{{\bf P}}(\varepsilon/2+\eta+\eta_1))} \\
\leq&
C_0\Big\{
\eta^2 \| H_E v\|_{L^2(U_{{\bf P}}(\varepsilon/2 + \eta_1))}
+ 
\!\!\!\!\!
\sum_{\beta \in {\mathbb N}^{3N},|\beta| < 2} \eta^{|\beta|}\| \partial^{\beta}
v\|_{L^2(U_{{\bf P}}(\varepsilon/2+\eta_1))}
\Big\}. \nonumber 
\end{align}
Furthermore, if the right hand side of \eqref{eq:dumt_nummer} is finite 
for all $\eta, \eta_1 > 0$ then 
$v \in W^{2,2}(U_{{\bf P}}(\varepsilon/2+\eta_1))$ for all $\eta_1 >0$.
\end{lemma}

\begin{proof}
Since $U_{{\bf P}}(\varepsilon/2 + \eta + \eta_1) \subset
U_{{\bf P}}(\varepsilon/2 + \eta_1)$, the estimate is obviously true
for $|\alpha_0|+|\alpha'| < 2$. Let $\alpha_0 \in {\mathbb N}^{3N}$,
$\alpha' \in {\mathbb N}^{3M}$ with
$|\alpha_0|+|\alpha'| = 2$. Choose $\phi \in C^{\infty}({\mathbb
R}^{3N})$, $0 \leq \phi \leq 1$, with $\phi \equiv 1$ on $U_{{\bf P}}(\varepsilon/2 + \eta +
\eta_1)$ and $\supp \phi \subset U_{{\bf P}}(\varepsilon/2 + \eta_1)$,
satisfying $\|\partial^{\gamma}\phi \|_{\infty} \leq
C_{\gamma} \eta^{-|\gamma|}$, with $C_{\gamma}$ independent
of $\eta, \eta_1$.

We can now estimate, using Lemma \ref{lem:obvious_domain} in
the third inequality below
\begin{align*}
\| \partial^{\alpha_0}&\partial^{\alpha'}_{x_{{\bf P}}}
v\|_{L^2(U_{{\bf P}}(\varepsilon/2+\eta+\eta_1))}
\leq \| \partial^{\alpha_0}\partial^{\alpha'}_{x_{{\bf P}}}
(\phi v)\|_{L^2({\mathbb R}^{3N})} \\
\leq & \, c \| (\phi v)\|_{W^{2,2}({\mathbb R}^{3N})} 
\leq C\left( \|H_E (\phi v) \|_{L^2({\mathbb R}^{3N})} +
\|\phi v \|_{L^2({\mathbb R}^{3N})} \right) \\
\leq & \,
C \Big\{ \|\phi H_E v \|_{L^2({\mathbb R}^{3N})} +
\|(\Delta \phi) v\|_{L^2({\mathbb R}^{3N})} \\
&
\,\,\,\,\,\,\,\,\,\,\,
+
2\|\nabla \phi\cdot  \nabla v \|_{L^2({\mathbb R}^{3N})} +
\|\phi v \|_{L^2({\mathbb R}^{3N})} \Big\} \\
\leq & \,
C \Big\{
\|H_E v \|_{L^2(U_{{\bf P}}(\varepsilon/2 + \eta_1))} \\
&
\,\,\,\,\,\,\,\,\,\,\,
+
c_1 \eta^{-1}\| \nabla v \|_{L^2(U_{{\bf P}}(\varepsilon/2 + \eta_1))}
  +
c_2 \eta^{-2}\| v \|_{L^2(U_{{\bf P}}(\varepsilon/2 + \eta_1))} \Big\},
\end{align*}
for some constants $c, C, c_1, c_2$.
Inequality \eqref{eq:dumt_nummer} follows by multiplying with
$\eta^2$.

The last statement of the lemma follows easily from
Lemma \ref{lem:obvious_domain}.
\end{proof}

Finally, we state and prove the properties of $V$ that we need in the proof of 
Lemma \ref{lem:diff_parallel}.

\begin{lemma}[Properties of $V$]
\label{lem:Properties_of_V}
Let $V$ be the Coulomb potential defined in \eqref{atomV}.
\begin{enumerate}
\item There exists $C_V >0$ such that for all $v \in
W^{1,2}({\mathbb R}^{3N})$ we have
\begin{equation}
\label{eq:Hardy}
\| (V-E) v \|_{L^{2}({\mathbb R}^{3N})} \leq C_V \| v
\|_{W^{1,2}({\mathbb R}^{3N})}.
\end{equation}
\item There exists a constant $L_V > 0$ (depending on
$\varepsilon$) such that for all $\alpha \in {\mathbb N}^{3M}$
with $|\alpha| \geq 1$, we have
\begin{equation}
\label{eq:analyticity_V}
\| \partial_{x_{{\bf P}}}^{\alpha} V \|_{L^{\infty}(U_{{\bf
P}}(\varepsilon/2))}
\leq L_V^{|\alpha|+1} |\alpha|! \,.
\end{equation}
\end{enumerate}
\end{lemma}

\begin{remark}
These are the only properties of $V$ that we need (together with Lemmas 
\ref{lem:obvious_domain} and \ref{lem:a_priori}). 
They are easily seen to hold (see argument in proof below) 
for potentials satisfying the general conditions in 
Remark \ref{rem:generalise}.
\end{remark}
 
\begin{proof}
The first property \eqref{eq:Hardy} is a consequence of
Hardy's inequality (see for instance \cite[Vol. II, p. 169]{ReSI}). 
To prove the second property, \eqref{eq:analyticity_V},
let $P_s$ be one of the index sets defined in Lemma \ref{lem:diff_parallel}. 
Notice that
$$
\partial^{\alpha}_{x_{P_s}} |x_j|^{-1} = 
\begin{cases} 
        0 & \text{ for  } j \notin P_s \\
        |P_s|^{-|\alpha|/2}\,\partial^{\alpha}_{x}|x|^{-1}
        \Big{|}_{x=x_j} 
        & \text{ for  } j \in P_s,
\end{cases}
$$ 
and
$$
\partial^{\alpha}_{x_{P_s}} |x_j-x_k|^{-1} = 
\begin{cases} 
        0 & \text{for  } j,k \notin P_s \text{ or } j,k \in P_s\\
        |P_s|^{-|\alpha|/2}\,\partial^{\alpha}_{x}|x|^{-1} \Big{|}_{x=x_j-x_k} & 
        \text{for  } j \in P_s, k \notin P_s.
\end{cases}
$$ 
Therefore, \eqref{eq:analyticity_V}
follows from the structure of $V$,  
the real analyticity of $x \mapsto |x|^{-1}$ away from $0$ 
and the definitions of 
$U_{{\bf P}}(\varepsilon/2)$ and $\partial^{\alpha}_{x_{\bf P}}$.
\end{proof}

\begin{proof}[Proof of Lemma \ref{lem:diff_parallel}]
Notice
that \eqref{eq:analyticity_V} trivially implies that for $j,
\eta >0$, $j \eta < 1$, $|\alpha| \geq 1$ we have
\begin{equation}
\label{eq:analyticity_useful_V}
\eta^{|\alpha|} \| \partial_{x_{{\bf P}}}^{\alpha} V
\|_{L^{\infty}(U_{{\bf P}}(\varepsilon/2+j\eta))}
\leq L_V^{|\alpha|+1} |\alpha|! j^{-|\alpha|} .
\end{equation}

We will prove that there exists $L_{\psi} > 0$, such that for
all $\eta \in (0,1)$ and all $j \in {\mathbb N}$ with $j
\eta <1$ we have, for all $\alpha \in {\mathbb N}^{3M}$,
$\alpha_0 \in {\mathbb N}^{3N}$, $|\alpha_0| \leq 2$,
$|\alpha|+ |\alpha_0|< 2+j$,
\begin{equation}
\label{eq:Induction}
\eta^{|\alpha|+|\alpha_0|} \| \partial^{\alpha_0}
\partial_{x_{{\bf P}}}^{\alpha}
\psi
\|_{L^{2}(U_{{\bf P}}(\varepsilon/2+j\eta))}
\leq L_{\psi}^{|\alpha|+|\alpha_0|+1} .
\end{equation}
Before proving \eqref{eq:Induction}, let us
note that Lemma \ref{lem:diff_parallel} follows easily
from it. In fact, let $\alpha \in {\mathbb N}^{3M}$,
$|\alpha| \geq 1$ and choose $|\alpha_0|=0$, $\eta = \varepsilon/(2 |\alpha|)$,
$j = |\alpha|$. Then \eqref{eq:Induction} becomes
$$
\| \partial_{x_{{\bf P}}}^{\alpha} \psi
\|_{L^{2}(U_{{\bf P}}(\varepsilon))}
\leq
L_{\psi}^{|\alpha|+1} (2/\varepsilon)^{|\alpha|}
|\alpha|^{|\alpha|} ,
$$
which is the statement of Lemma \ref{lem:diff_parallel}.

We now prove \eqref{eq:Induction} by
induction in $j$. For $j=0,1$, there is nothing to prove
since we know that $\psi \in W^{2,2}({\mathbb R}^{3N})$. 
Let $L_{\psi}$ be sufficiently large
for \eqref{eq:Induction} to be true for $j=0,1$ and
satisfying furthermore,
\begin{equation}
\label{eq:choice_L}
L_{\psi} \geq \max \Big\{2 L_V,C_0(1 + \sum_{|\beta| < 2} 1) \Big\}.
\end{equation}
Here the sum is over all $\beta \in {\mathbb N}^{3N}$ with $|\beta| <
2$, $C_0$ is the constant from Lemma \ref{lem:a_priori},
and $L_V$ is the constant from \eqref{eq:analyticity_V}.

Suppose that we have proved \eqref{eq:Induction} for all
$j \leq j_0$ and all $\eta \in (0,1)$ with $\eta j < 1$. We
will prove that
\eqref{eq:Induction} holds for $j=j_0+1$ and all $\eta \in
(0,1)$ with $(j_0+1)\eta <1$.\\
Let $|\alpha|+|\alpha_0|< 2 + j_0$. Then clearly 
$U_{{\bf P}}(\varepsilon/2+(j_0+1)\eta) \subset 
U_{{\bf P}}(\varepsilon/2+j_0 \eta)$. Therefore,
$$
\eta^{|\alpha|+|\alpha_0|} \| \partial^{\alpha_0}
\partial_{x_{{\bf P}}}^{\alpha}
\psi
\|_{L^{2}(U_{{\bf P}}(\varepsilon/2+(j_0+1)\eta))}
\leq
\eta^{|\alpha|+|\alpha_0|} \| \partial^{\alpha_0}
\partial_{x_{{\bf P}}}^{\alpha}
\psi
\|_{L^{2}(U_{{\bf P}}(\varepsilon/2+j_0\eta))},
$$
and the result holds by the induction hypothesis. So we only
have to consider the case $|\alpha|+|\alpha_0| = 2 + j_0$.
Choose a decomposition $\alpha = \alpha' + \alpha''$, with
$|\alpha'|=2-|\alpha_0|$, i.e. with $|\alpha''|=j_0$. 
Using Lemma \ref{lem:a_priori}
with $\eta_1 = j_0 \eta$, $v=\partial_{x_{{\bf P}}}^{\alpha''} \psi$
we find
\begin{multline}
\label{eq:a_priori_induction}
\eta^{2+j_0} \| \partial^{\alpha_0}
\partial_{x_{{\bf P}}}^{\alpha} \psi
\|_{L^{2}(U_{{\bf P}}(\varepsilon/2+(j_0+1)\eta))} \\
\leq
C_0\Big\{
\eta^{2+j_0} \| H_E \partial_{x_{{\bf P}}}^{\alpha''} \psi
\|_{L^{2}(U_{{\bf P}}(\varepsilon/2+j_0\eta))}  \\
+
\sum_{|\beta| < 2}
\eta^{|\beta|+j_0} \|
\partial^{\beta} \partial_{x_{{\bf P}}}^{\alpha''} \psi
\|_{L^{2}(U_{{\bf P}}(\varepsilon/2+j_0\eta))} \Big\}. 
\end{multline}
Since $H_E \psi = 0$, we get
\begin{align}
\label{eq:energy_term}
&\eta^{2+j_0} \| H_E \partial_{x_{{\bf P}}}^{\alpha''} \psi
\|_{L^{2}(U_{{\bf P}}(\varepsilon/2+j_0\eta))} \nonumber \\
&=
\eta^{2+j_0} \Big\|
\sum_{\gamma < \alpha'', \beta + \gamma = \alpha''}
\binom{\alpha''}{\gamma}
\left( \partial_{x_{{\bf P}}}^{\beta}(V-E) \right)
\partial_{x_{{\bf P}}}^{\gamma} \psi
\Big\|_{L^{2}(U_{{\bf P}}(\varepsilon/2+j_0\eta))}
\end{align}
We now use \eqref{eq:analyticity_useful_V}, the
combinatorical result from Proposition \ref{prop:combinatorics}
together with the induction hypothesis,
to estimate \eqref{eq:energy_term} as
\begin{eqnarray}
\label{eq:highest_derivatives}
&&\eta^{2+j_0} \| H_E \partial_{x_{{\bf P}}}^{\alpha''} \psi
\|_{L^{2}(U_{{\bf P}}(\varepsilon/2+j_0\eta))} \nonumber \\
&\leq&
\sum_{k=1}^{|\alpha''|} \binom{|\alpha''|}{k}
L_V^{k+1} k! j_0^{-k} L_{\psi}^{|\alpha''|-k+1}
\leq \sum_{k=1}^{|\alpha''|} L_V^{k+1}
L_{\psi}^{|\alpha''|-k+1} \nonumber \\
&\leq&
L_{\psi}^{|\alpha''|+2} \sum_{k=1}^{|\alpha''|}
(L_V/L_{\psi})^{k+1} \leq L_{\psi}^{|\alpha''|+2}.
\end{eqnarray}
Here we used the assumption $L_{\psi} \geq 2 L_V$ from 
\eqref{eq:choice_L} in the last estimate.

Due to the induction hypothesis we can also estimate the other term in
\eqref{eq:a_priori_induction}
\begin{eqnarray}
\label{eq:lower_derivatives}
\sum_{|\beta| < 2}
\eta^{|\beta|+j_0} \|
\partial^{\beta} \partial_{x_{{\bf P}}}^{\alpha''} \psi
\|_{L^{2}(U_{{\bf P}}(\varepsilon/2+j_0\eta))}
&\leq &
\sum_{|\beta| < 2} L_{\psi}^{|\alpha''|+|\beta|+1}.
\end{eqnarray}
So using \eqref{eq:highest_derivatives} and
\eqref{eq:lower_derivatives}, we can estimate
\eqref{eq:a_priori_induction} as
$$
\eta^{2+j_0} \| \partial^{\alpha_0}
\partial_{x_{{\bf P}}}^{\alpha}
\psi
\|_{L^{2}(U_{{\bf P}}(\varepsilon/2+(j_0+1)\eta))}
\leq
L_{\psi}^{|\alpha|+|\alpha_0|+1}\left( \frac{C_0(1+\sum_{|\beta|< 2}
1)}{L_{\psi}} \right).
$$
The last factor is $\leq 1$, by the choice of $L_{\psi}$ 
(see \eqref{eq:choice_L}).
That finishes the proof of \eqref{eq:Induction} and
therefore of Lemma \ref{lem:diff_parallel}.
\end{proof}

\begin{acknowledgement}
All four authors thank the organizers of the program {\it Partial Differential Equations and Spectral Theory} for invitations
to the Mittag Leffler institute where part of the work was done.
Furthermore, parts of this work has been carried out at various institutions, whose hospitality is gratefully acknowledged: Aalborg University (SF, MHO, THO), The Erwin Schr\"{o}dinger Institute (T\O S) and Universit\'{e} Paris-Sud (T\O S).
Financial support from the Danish Natural Science Research Council, 
European Science Foundation Programme {\it Spectral Theory and Partial Differential Equations} (SPECT), and EU IHP network
{\it Postdoctoral Training Program in Mathematical Analysis of
Large Quantum Systems},
contract no.\
HPRN-CT-2002-00277 is
gratefully acknowledged.\\
SF was supported by a grant from the Carlsberg
Foundation.\\
Finally, SF wishes to thank B. Helffer and L. T'Jo\"{e}n for useful discussions and encouragement. 
\end{acknowledgement}

%
       \bibliographystyle{abbrv}
        \bibliography{ref}

\end{document}